\documentclass[12pt]{article}
\newcommand{\arcsec}{$^{\prime\prime}$}
\newcommand{\msun}{$M_{\odot}$}
\usepackage{graphicx}
\begin{document}
{\LARGE \bf Quasar lensing}

{\bf Introduction}

Strong gravitational lensing is a phenomenon which occurs when the lines of sight
to a foreground and background object nearly coincide, resulting in multiple imaging
of the background object. Although quite rare, it offers an important diagnostic of
masses and mass distributions in foreground objects ranging from stars to clusters of 
galaxies, and has the important advantage of being sensitive to all kinds of matter, both 
baryonic and dark. Strong lensing also gives magnified views of background objects,
allowing easy study of otherwise inaccessible quantities, and potentially cosmological
information owing to its sensitivity to combinations of mass density and lengths within
the universe.

Quasars are relatively rare phenomena, and lensed quasars, in which a foreground galaxy
provides the gravitational deflection in the right place, are correspondingly rare. 
However, they have some unique advantages: quasars allow easy access to the high
redshift universe; quasars are bright, allowing easy study of subtle effects and 
potentially allowing complete samples to be built; quasars are variable, allowing
cosmological information to be derived from time delays; quasars emit at multiple
wavelengths, allowing detailed study of propagation effects. This review gives an overview
of the history of quasar lensing, and a summary of its main applications. The applications
fall into three parts: the use of quasars as probes of lens galaxies, both their
stellar content via microlensing and their dark matter content via the fitting of 
models to lensed images; the use of quasar lenses to probe the structure and properties 
of the quasars themselves; and the use of quasar lenses for cosmology. A number of previous 
reviews have addressed some or all of these issues; see, for example Wambsganss (1998), 
Claeskens \& Surdej (2002), Courbin, Saha \& Schechter (2002), Kochanek \& Schechter 
(2004), Kochanek (2004), Wambsganss (2004), Jackson (2007), Zackrisson \& Riehm (2010),
Bartelmann (2010), Schmidt \& Wambsganss (2010). In this review individual theoretical 
results will be 
presented as needed, without derivation; the interested reader can refer to the standard 
text by Schneider, Ehlers \& Falco (1992) for more detail. Finally I outline the possible 
future applications of quasar lensing, and the observational programmes which will develop
the subject in the coming years.

{\bf Historical introduction}

Lensing remained an interesting theoretical possibility for most of the twentieth 
century. The deflection of light by a mass was calculated classically by 
Soldner (1801) and correctly by Einstein as a consequence of general relativity, and
famously observed by Eddington in 1919 using position shifts of stars close to the
Sun during a solar eclipse. The possibility of multiple images formed by
individual objects was considered by Chwolson (1924). However, it was not until
1979 that the first gravitational lens system was actually found.

The first lens system, Q0957+561, was discovered by Walsh et al. (1979) during the 
course of optical followup of sources found in a radio survey at 966~MHz with the 
Jodrell Bank telescope. The source object in this system is a radio-loud quasar at 
redshift 1.41, doubly imaged by a galaxy at redshift 0.36 (Fig. 1). The large (6\arcsec) 
separation of the two images in this system is due to the fact that the lensing galaxy 
is assisted by a cluster at the same redshift; this allowed Walsh et al. to measure 
separately the redshifts of the two lensed images, whose spectral similarity
supported the hypothesis that their light originates from the same background object.
In the next few years, further objects were discovered, many of them in radio surveys;
these have the advantage that the sources they contain are predominantly non-thermal
emitters such as quasars, without contamination from stellar processes.
A typical non-thermal radio source consists of a central bright flat-spectrum core, 
corresponding to the active nucleus at the centre of the host galaxy, and extended
steep-spectrum emission in lobes which result from ejection of material from the active 
centre. Although the Q0957+561 system has a double image of the core, further quasar lenses
were found in which the extended radio emission was gravitationally imaged, resulting
in rings (e.g. MG1131+0456, Hewitt et al. 1988).

\begin{figure}
\includegraphics{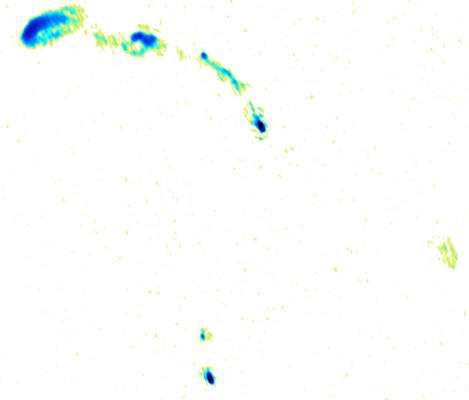}
\caption{A modern view of the Q0957+561 lens system, made with the e-MERLIN telescope
at 5GHz. The quasar consists of a core, radio jet (top left), and lobe (bottom right). 
The second image of the core (bottom of the image) is close to the lensing galaxy. 
(Image credit: Muxlow, Beswick \& Richards, Manchester).}
\end{figure}

Searches for lensed radio sources divided into two main areas. The first, following the
success of the MG survey, targeted extended radio lobes, looking for rings associated with 
the presence of galaxies in front of them (Leh\'ar et al. 2001, Haarsma et al. 2005). 
The second method involved systematic targeting of flat-spectrum radio sources, in 
which the central bright point component dominates the radio emission. This makes 
lensing relatively easy to recognise, although the low optical depth to lensing means 
that a large number of candidates must be examined to find relatively few lenses. The 
CLASS survey (Myers et al. 2003, Browne et al. 2003) is still the largest systematic 
radio survey, and produced 22 lens systems from a parent sample of 16503 northern objects 
initially observed with the VLA and followed up at higher resolution using Merlin and 
VLBI. A southern extension of this survey also exists (e.g. Winn et al. 2001) 
which discovered a further 4 lens systems. Most of the lenses are elliptical galaxies, 
since these are generally more massive and hence dominate the lensing cross-section, but 
a significant minority
are later types. Subsamples of these 22 are still important for many of the
astrophysical and cosmological applications that will be discussed later in this review.

About 90\% of quasars do not have bright radio emission. Searches for lensed radio-quiet
quasars began soon after the Walsh et al. (1979) initial discovery, most successfully
using the HST. Lensed optical quasars began to be discovered in substantial numbers 
following the availability of the Sloan Digital Sky Survey (SDSS, York et al. 2000) and 
subsequently of quasar catalogues derived from it. The Sloan Quasar Lens Search (SQLS; 
Oguri et al. 2006, Inada et al. 2008, Inada et al. 2010, Inada et al. 2012) has discovered
% 2006: AJ132 999   2008: 135 496  2010:AJ 140 403 2012:143 119
the majority of these, by following up quasars which show extended emission on SDSS
images, suggesting the presence of secondary lensed images, or of a lensing galaxy, or
both. The SQLS has produced about 30 quasar lenses in the most recent catalogue, including
some of the widest-separation lenses known (Inada et al. 2003, Inada et al. 2006). In 
addition
% 2006:apj 653 l97
a number of smaller surveys have increased the number of lenses, using various methods.
These include the use of higher-image quality supplementary surveys such as UKIDSS
(Lawrence et al. 2007) to find small-separation or high-flux-ratio lenses (Jackson et
al. 2009, 2012).

The most complete current database, the Masterlens project (Moustakas et al. 2012), 
lists 120 quasar lenses of
which 71 result from two surveys (SQLS and CLASS) and the remainder are serendipitous
discoveries or result from smaller surveys. It is this sample that forms the current
basis of the scientific results discussed in this review. Future instruments will
increase this sample by many orders of magnitude, allowing correspondingly more detailed
and wide-ranging science which is discussed in the final section.

{\bf QL as probe of lens galaxies}

{\it Introduction}

Any lens system allows the projected mass distribution of the lens to be probed. Since
the lens is typically an elliptical galaxy at significant ($\sim$0.5) redshift, this
is an intrinsically interesting observation, as detailed study of stellar dynamics at 
this distance is a painful operation involving large telescopes and long integration 
times. Moreover, lensing gives the projected mass of the lens within the Einstein 
radius with unique accuracy. There is less good news, however; the probes of
the lens potential are obtained only at points where an image is formed, and typically
these are limited because the emission from quasars is usually very compact. The 
resulting problem of mass reconstruction is thus typically underconstrained, giving 
a serious problem with degeneracies when trying to deduce other mass properties of 
the lens, particularly the radial mass profile (e.g. Saha 2000, Wucknitz 2002, 
Kochanek 2002, Kochanek 2004). Probably the most depressing degeneracy is the so-called 
mass sheet degeneracy (Falco et al. 1985, Gorenstein et al. 1988), which results from 
the observation that it is possible for an intervening mass distribution to leave the 
lensed image positions and fluxes undisturbed, while rescaling the overall potentials 
and time delays together with the (unobservable) 
source position. This is a potentially serious problem for cosmological investigations.

One approach which alleviates some of the degeneracy problems is to abandon the use of 
point-source lenses in favour of lensed galaxies, which have extended source structure 
and thus give constraints at multiple radii in the image plane. This has had considerable 
success in  the investigation both of overall mass profiles of galaxies and of sub-galactic 
level substructure. The most significant work in this direction stems from the SLACS survey 
(Bolton et al. 2006, Koopmans et al. 2006, Bolton et al. 2008). A second approach is to 
concentrate on a small sample of high-value objects and do the followup observations which 
are necessary to break the degeneracies; this is mainly to derive additional constraints 
on the mass profile from observations in regimes where extended sources are visible in 
addition to the point-like quasar. This approach also requires extensive investigation 
of the foreground (Fassnacht et al. 2006, Momcheva et al. 2006), in order to derive the 
information about nearby objects required to ease the mass-sheet degeneracy. 
(Alternatively, very rare objects, such as lenses with two different sources at 
different redshifts, can be used to determine the source position and thus break the 
MSD completely). The third approach is to use quantities derived from the observations 
which are sensitive to a limited subset of the mass properties of the lens. Both the 
second and third approaches have been used for quasar lenses.

{\it Lensed substructure}

The main attraction of quasar lenses is that they provide a probe of sub-galactic-scale
matter structure, which is in turn relevant to a strong prediction of CDM models of 
structure formation. In
such scenarios, structure in the Universe forms in a hierarchical way, with small dark
matter clumps coalescing into larger haloes as time passes to form steadily larger 
agglomerations (White \& Rees 1978). Baryons lose potential energy by non-gravitational 
means and therefore settle in to the resulting potential wells, forming galaxies, groups 
and clusters. Because many physical processes are involved, how this happens is quite 
complicated, and in practice semi-empirical recipes are used for describing the
process (Blumenthal et al. 1986, Ryden 1988, Gnedin et al. 2004). There are also ongoing 
processes which may rearrange the baryonic material in galaxy-sized haloes, including the 
influence of supernovae in lower-mass haloes (Heckman et al. 1990) and of periodic 
ejections of matter from active nuclei in the centres (e.g. Begelman et al. 1991, Croton 
et al. 2006); these processes are collectively known as feedback. Complications associated
with baryon physics cannot be avoided in lensing systems, because typical gravitationally
lensed images form at radii of about an arcsecond, corresponding to 5-10kpc in projection
against the lens galaxy. At this radius, dark matter is expected to contribute at the
level of a few tens of percent to the projected matter distribution, and baryon processes
are therefore dominant.

The substructure debate matters, because CDM works so well on cluster and supercluster 
scales that its predictions on smaller scales are one of its few possible failure modes. 
In our own Galaxy, the observation that fewer luminous satellites were found than CDM
subhalo models would predict (Moore et al. 1999, Klypin et al. 1999), generated a vast
literature reflecting the importance of the problem. Possible ways out of the problem 
included finding at least some of the missing satellites (Belokurov et al. 2006, 2007; 
Zucker et al. 2006a,b), 
% zucker apj 643 L103, 650 L41
% belok apj 647 l111, 654 897
or finding some way in which they might be present but not accrete gas and form stars 
(Bullock et al. 2000). The current situation is that it is difficult to explain both 
the incidence of substructure and its dynamical properties and stellar content (e.g. 
Boylan-Kolchin, Bullock \& Kaplinghat 2011, Boylan-Kolchin, Bullock \& Kaplinghat 2012) 
although possibilities exist including detailed adjustments corresponding to detailed 
treatments of the physics, or gross changes such as an alteration in the overall mass 
of the Milky Way and thus of its expected subhalo content. Curiously, despite the apparent 
lack of substructure on sub-Magellanic Cloud mass scales, the presence of substructure
on scales as massive as the Magellanic Clouds themselves is mildly anomalous, unless 
the Milky Way has a mass towards the upper end of the allowed range (Boylan-Kolchin et 
al. 2010).

{\it Lensed substructure in other galaxies}

Relief from detailed arguments about CDM substructure in our own Galaxy can be had by
considering substructure probes in other galaxies, in which details can be swept under
the observational carpet - or less cynically, a larger number of objects can be probed in 
less detail, thus accounting for the possibility that our Galaxy may be untypical. The 
main mass probe in other galaxies at cosmological distance is gravitational lensing.

The observation of a gravitational lens system yields a set of observed positions and
flux densities of lensed images. As already discussed, this set of observables does not 
give unique information about the mass distribution, and in general a large number of 
macromodels (a term generally used for overall mass distributions, or at any rate for 
components of spatial frequency in mass distribution of a few kpc or larger) are 
compatible with images of a single object. Some plausibility arguments can be used to 
restrict the available set of macromodels, mainly the observation that well-constrained 
lens systems, with stellar dynamics and extended images, suggest approximately 
isothermal\footnote{This corresponds to a surface density profile $\Sigma\propto r^{-1}$, 
or a 3-dimensional density profile $\rho\propto r^{-2}$.} distributions of mass 
(e.g. Cohn et al. 2001, Rusin et al. 2003, Koopmans et al. 2006). This density 
profile corresponds to a flat rotation curve. On top of the macromodel, any 
smaller-scale perturbations will affect the image positions and fluxes. Since image 
positions depend on the first derivative of the projected potential distribution and 
fluxes on the second, smaller perturbations are expected to be detectable in image fluxes.

A number of relations between fluxes of individual lensed images exist which are 
independent of, or at least relatively insensitive to, the details of the macromodel. For 
example, ``cusp'' configuration lens systems, in which the source lies close to the cusp 
of the astroid caustic produced by the lens galaxy (Fig. 2), produce three bright images 
close together on the opposite side of the Einstein ring from a single faint image. The 
brightness of the central image of these three should be equal to the combined brightness 
of the outer two (Schneider \& Weiss 1992; see also Keeton, Gaudi \& Petters 2003,2005; 
Congdon, Keeton \& Nordgren 2008 for more detailed treatment of other cases), and any 
departure from this indicates a non-smooth mass model. The relation holds because the
images form at a very similar, and relatively flat, part of the Fermat 
surface\footnote{The Fermat surface is a very useful way of thinking about gravitational 
lens optics. Imagine a source, viewed by the observer in projection on to the lens plane, 
with contours drawn according to the light travel time of rays originating in the source, 
bending in the lens plane and reaching the observer. These contours are simply concentric 
circles centred on the source, with a central stationary point (a minimum) at which 
Fermat's principle dictates the formation of an image. If we then introduce a galaxy, 
which distorts these contours, 
we eventually reach a point at which further stationary points (a maximum and saddle point)
simultaneously form.} in which unphysically large changes in the macromodel would be 
needed to produce disagreements -- known in the literature as ``cusp violations'' -- 
with the expected relation. On the other hand, small-scale structure can produce cusp 
violations relatively easily. Constraining small-scale structure is in principle a matter 
of counting the number and magnitude of cusp violations in a sample of quasar lenses.

\begin{figure}
\begin{tabular}{cc}
\includegraphics[width=6cm]{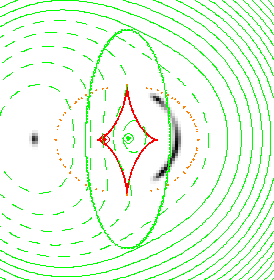}&
\includegraphics[width=5.7cm]{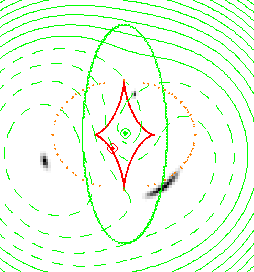}\\
\end{tabular}
\caption{Geometry of cusp (left) and fold (right) lenses. In each case the galaxy is
centred on the green dot, and the projected position of the source is the red dot. 
The astroid caustic (the red diamond-shaped feature) separates 2-image from 4-image
systems. The greyscale represents the observed images of the source, which form at
positions which are stationary in the time-delay surface (indicated by green contours).}
\end{figure}

There are a number of reasons why the problem is not that easy. The first is that anomalous
fluxes can be produced not only by CDM substructure on 10$^6$-10$^9$\msun scales, but
also by the movements of individual stars in the lensing galaxy which create a caustic
pattern of differential magnification which tracks across the field in timescales of years,
with individual events happening on shorter timescales. This phenomenon, microlensing, 
described in detail later, is itself extraordinarily interesting, but a contamination
for the current purpose. It can be got around by using sources which have large sizes
relative to the scale of the microlensing caustic pattern, which in practice is about 
1~$\mu$as. The cores and VLBI-scale jets of radio sources, with a typical intrinsic 
angular size\footnote{The size of a compact radio source
is controlled by where the optical depth to synchrotron self-absorption
becomes 1. This is typically about 1~mas for a source of around 1~Jy, although it
becomes smaller with increasing frequency, and it decreases as the square root of the flux.
Sources typically found by the Square Kilometre Array, which will be sensitive to
sources of 1~$\mu$Jy, may therefore show radio microlensing.} of about 1~mas, fulfil 
this condition, but optical quasars do not. The second problem, even for radio quasars,
concerns propagation effects. Radio waves can be scattered while propagating through 
ionized media; the details are complicated (Rickett 1977) but the effect is to produce 
flux variations, which can be seen on timescales of weeks to years if the scattering
screen is in our own Galaxy. By definition, there is a possible source
of foreground screen in gravitational lens systems, in the form of the lensing galaxy,
as well as a nearer screen in our own galaxy. Koopmans et al. (2003) found evidence
for this in at least one object during a monitoring campaign of some radio lenses, but
it is likely to be a small effect compared with most of the observed flux anomalies. The
third possible problem is the effects of intrinsic variation of the quasar, coupled with
a differential time delay between the images. Even though time delays  and flux density 
variation are useful for measuring the Hubble constant, for the present purpose they are 
a nuisance. Again, however, the level of variation of most radio sources does not seem 
to be significant enough to be a major problem, and can be averaged out if observations
are made for periods much longer than the time delay. In the optical, extinction
is present, and can be used to probe the properties of the dust in the lensing galaxy
by using the fact that the same object's light path passes through two different
regions of the galaxy (El\'{\i}asd\'ottir et al. 2007).

The first obvious flux anomaly was pointed out by Mao \& Schneider (1998) in the CLASS
lens system B1422+231; this is a cusp system which produces a violation which requires 
a significant amount of substructure (about that predicted by CDM) in order to give a 
significant chance of reproducing the observed anomaly. Other examples of flux 
anomalies which defied smooth macromodels soon followed (Fassnacht et al. 1999,
Metcalf \& Zhao 2002, Chiba 2002, Saha, Williams \& Ferreras 2007), leading to the 
first attempt to address the overall statistics (Dalal \& Kochanek 2002, see also 
Kochanek \& Dalal 2004). Using seven four-image lenses, Dalal \& Kochanek derived an 
overall substructure contribution of 0.6-7\% (2$\sigma$ confidence) in substructures 
between 10$^6$\msun and 10$^9$\msun, in rough agreement with the overall predictions of 
CDM. However, this substructure appears to be in the wrong place (Mao et al. 2004); dark 
matter, and hence dark matter substructure, should in CDM models be less centrally 
concentrated than the baryons, and such levels of substructure at projected radii of 
5-10~kpc are thus surprisingly high -- a curious contrast to the ``missing satellite'' 
problem in our own Galaxy. Incidentally, the presence of a tension between lensing 
observations and CDM probably gives a severe problem for models involving significant 
amounts of warm dark matter (WDM), which would predict even less substructure (Miranda 
\& Macci\`o 2007).

A more sophisticated approach to CDM testing can be taken, if instead of calculating
an average contribution of substructure ``expected'' at the projected Einstein
radius, we instead take an actual CDM halo simulation and investigate its lensing
properties. Early attempts, with lower resolution simulations, produced mixed
results (Brada\v{c} et al. 2004, Macci\`o et al. 2006, Amara 2006), but generally confirmed
the picture of an excess of flux anomalies compared to the expected incidence in
$\Lambda$CDM. As better simulations became available, they have been used for these
comparisons (Xu et al. 2009, see also Chen, Koushiappas \& Zentner 2011 for more 
detailed treatment of halo-to-halo variations) using, for example, the Aquarius dark-matter 
simulations. There are a number of limitations with such investigations. The 
two main problems are that the simulations are being pushed to the limit of their 
resolution, since they are being asked questions about mass condensations on scales 
down to 10$^6$\msun, comparable to the lowest masses considered in the simulations, 
and that the effect of baryons in modifying the structure of the 
subhaloes is not taken into account. Until higher-resolution simulations with extra
physics are available, however, this is the best that can be done. Xu et al.'s 
conclusion was that the cusp violations in existing lenses clearly exceeded the level 
of violation which would be expected in the dark matter simulations. Two important 
caveats to this emerged in subsequent work, however. Firstly, detection of substructure 
along the line of sight means just that, and the substructures which produce the flux 
anomalies do not have to be within the lensing galaxy (Metcalf 2005a,b, Inoue \& Takahashi 
2012). Xu et al. (2012) suggested that 20-30\% of the substructure could be outside the 
lensing galaxy, somewhere along the line of sight. If correct, this would potentially 
alleviate the tension between lensing observations and CDM, although it is probably fair 
to say that more work, both theoretical and observational, is needed before this 
can be regarded as well established. Secondly, finite source-size effects may modify
the statistics of substructure detection (Dobler \& Keeton 2006; Metcalf \& Amara
2012).

The sample of seven radio-loud lenses used for substructure studies has remained 
largely unchanged in the last
decade, owing to the current difficulty of finding significant numbers of new radio
lenses with existing telescopes. There are a number of alternative approaches. 
The first involves the use of observational brute force; to target radio-quiet lenses, 
but observe flux densities in parts of the electromagnetic spectrum where the source 
has significantly greater size than the microlensing characteristic size of 1$\mu$as. 
The obvious choice is the mid infra-red, where the source is expected to consist of a 
more extended thermal component than the accretion disk which radiates in the optical 
and ultraviolet. Despite the difficulties of observing in this waveband, a number of 
successful programmes have been carried out (Chiba et al. 2005, Fadely \& Keeton 2011;
Fadely \& Keeton 2012) resulting in the detection of a number of other flux anomalies,
and measurement of their likely masses. These range from the 10$^{7.3}$\msun and
10$^{7.7}$\msun clumps found in MG0414+0534 and HE0435$-$1223 respectively (MacLeod et 
al. 2012, Fadely \& Keeton 2012) to larger perturbations (10$^9$\msun in SDSS~J1029+2623
and 2$\times 10^8$\msun in 1938+666; Kratzer et al. 2011, Vegetti et al. 2012).
By contrast, the substructure identified in galaxy-galaxy lenses is often larger 
(e.g. Vegetti et al. 2010). Radio-quiet quasars are also not radio-silent, and flux 
densities have been measured for a number of such lens systems (Kratzer et al. 2011, 
Jackson 2011). Indeed, one would expect that all quasars emit measurable flux density 
at radio frequencies (White et al. 2007) with current instruments such as the EVLA and 
e-MERLIN.

% XXX some more on previous paragraph

An alternative approach is to use the presence of additional observational constraints,
such as those provided by radio jets in quasars, to give additional observational 
constraints, in the case where the jets can be detected in more than one lensed image. 
This was first attempted by Metcalf (2002) in the case of the lens system CLASS~B1152+199 
(Myers et al. 1999, Rusin et al. 2002) and a detection was claimed in this case. With 
further investigations using VLBI, other lenses have been shown to require substructure 
(MacLeod et al. 2012) and this may be a promising path to more detailed substructure 
measurements in the future, given sensitive VLBI observations and high resolution 
(Zackrisson et al. 2012). Currently, one of the most puzzling cases is the four-image 
lens system CLASS~B0128+437 (Phillips et al. 2000), in which the source consists of three 
radio components separated by a few milliarcseconds and resolvable with VLBI. Attempts 
to fit the positions of the twelve resulting images fail badly (Biggs et al. 2004). In 
this object, also, the SIE macromodel which properly fits the four images on arcsecond 
scales contains an implausibly large amount of external shear, which is inconsistent 
with the observed number of surrounding galaxies, and also does not fit the extended 
structure around the images seen by adaptive optics observations (Lagattuta et al. 2010). 
A fruitful area of future investigation may well be to try and combine the flux and 
astrometric anomalies in a sample of lenses; although in the case of astrometric 
anomalies, unlike flux anomalies, the anomaly is always likely to be underestimated 
since it can be absorbed into the macromodel (Chen et al. 2007). An alternative issue 
for the future is the proposal that time delay measurements may also be useful for 
measuring the effects of substructure, which can in extreme cases change the sign of 
the differential time delay between two images (Keeton \& Moustakas 2009).

Having detected mass substructures, we can ask whether they consist purely of dark matter
or whether they contain stars. In many cases, flux anomalies can be explained by a
mass contribution from a substructure which corresponds to an observed luminous satellite 
galaxy (Schechter \& Moore 1993, Ros et al. 2000, McKean et al. 2007, Macleod, Kochanek \& 
Agol 2009), although in some cases (McKean et al. 2007) the mass model of the satellite is
contrived, indicating that further mass structures may be needed. The number of bright
subsidiary deflectors may be larger than expected from simulations (Shin \& Evans 2008), 
a problem which may be resolvable if some of them are actually line-of-sight structures,
or explicable as a selection effect if brighter condensations are rendered more effective 
at causing flux anomalies because they have higher central densities.

{\it Black holes and central potentials of lens galaxies}

A further important astrophysical application of quasar lens systems - and in particular,
of radio quasar lens systems - is the detection and study of ``odd'' images. All 
gravitational lens systems have an odd number of images, usually 3 or 5, resulting
from the properties of the lens Fermat surface. One 
of these images is always a Fermat maximum which forms very close to the centre of the 
lens galaxy. For most realistic mass distributions, this maximum in the surface is very 
sharp, which implies that the corresponding image is very faint (Wallington \& Narayan 
1993, Rusin \& Ma 2001, Keeton 2003). How faint it is depends on the geometry of the
lens system and the degree to which the central potential is singular; if the potential
is dominated by a massive black hole, the corresponding image can be hugely 
demagnified and for all practical purposes invisible. The geometry of the lens system
has an effect because this determines the separation of the central image from the
centre of the lensing galaxy. Three-image lens systems, particularly those with
high primary-secondary flux ratios, create central images further from the lens 
centre and which are thus less demagnified. Five-image systems, with four bright 
images, are expected to nearly always contain an undetectably faint fifth image because 
the symmetry of the lens configuration places it close to the galaxy centre.

Propagation effects are likely to be particularly acute when trying to detect central 
images, because the light path passes straight through the lens galaxy centre where the 
concentration of dust and ionized gas is high. The use of radio lenses, where the 
galaxy is unlikely to be visible, is required, but relies on the expectation or hope 
that the central image will not be scattered out of existence. Nevertheless, detection 
of a faint radio image is not the end of the story, since it may result from low-level 
radio emission from the core of the lensing galaxy. In principle this
can be distinguished by observation at different frequencies and examination of the
radio spectrum to see if it differs from the other images.

Observations to detect odd images are very difficult, because they require a combination 
of high resolution and extreme radio sensitivity. A comprehensive theoretical study 
(Keeton 2003) of likely lens mass profiles, based upon HST observations of Virgo 
ellipticals (Faber et al. 1997) showed that central image detections were likely only 
once flux density levels of 10-100~$\mu$Jy were reached. This level is only now becoming 
routine thanks to
high-bandwidth upgrades to the VLA (now the JVLA) and MERLIN (now e-MERLIN). With 
older instruments, there is only one secure detection of a central image in a galaxy-mass
lensing system, namely PMNJ~1632$-$0033 (Winn et al. 2002, 2003, 2004), although other
systems in which a softer cluster potential is the primary deflector have shown central
images (SDSS1004+4112, Inada et al. 2005). In other cases, considerable effort with older
instruments has yielded only upper limits (Boyce et al. 2006, Zhang et al. 2007).

The central image in PMNJ~1632$-$0033 implies two limits; on the mass of the central
black hole (which must be less than 2$\times 10^8 M_{\odot}$) and a lower limit on the 
central surface mass density. These can alternately be rewritten as joint limits on 
the index of the central mass power law and black hole mass. In principle, the 
degeneracy between these two parameters can be broken in the case where the third 
image can itself be split by the combined lensing effect of the black hole and central 
stellar cusp into two images (Mao, Witt \& Koopmans 2001).
Such a detection, although very much more difficult and requiring another factor of
10 in sensitivity, would be very exciting because it would enable the immediate 
measurement of the black hole mass and central stellar cusp density separately. Even
the detection of third images, or significant limits thereon, in a number of radio
lenses would give a powerful indication of the evolution of the central regions of
elliptical galaxies between $z=0.5$ and the present day.

{\bf QL microlensing: a probe of quasar structure}

The combined effect of many stars within the lensing galaxy is to produce a maze of
caustics, elongated regions of high magnification with dimensions of microarcseconds
which form an intricate pattern across which the source moves. Sources with angular
sizes smaller than the characteristic scales of this pattern suffer time-dependent
magnification as the pattern moves across them, and consequently the brightness
of each lensed image varies as its line of sight crosses the caustic pattern. The 
details of the resulting effects on the image lightcurves were calculated in the years 
following the discovery of Q0957+561 (Chang \& Refsdal 1979, Paczynski 1986, Kayser et 
al. 1986, Kayser \& Refsdal 1989).
% pacz86 paper is apj 301 503
It was first observed by Irwin et al. (1989) in the lens system Q2237+0305 (the 
``Einstein cross'', Huchra et al. 1985) which is a four-image lensed system produced 
by a low-redshift spiral galaxy with a high central stellar density around the
lensed images; the system is also useful because the time delays are small, much
less than the timescale of variations due to microlensing.

The most basic information carried by the microlensing lightcurves is a combination
of source size and mass of the microlensing objects (Schmidt \& Wambsganss 1998,
Wyithe et al. 2000, Yonehara 2001, Kochanek 2004). However, the fact that sources 
of different sizes respond differently to microlensing by the stars in the lens galaxy 
offers an opportunity to study sources in great detail (Wambsganss \& Paczynski 1991), 
as well as a way to infer the presence of microlensing by differences in spectra between
one image and another (e.g. Wisotzki et al. 2003).

The central region of quasars contain an accretion disk close to the central supermassive
black hole, with temperatures of over 10000~K and producing hard UV and X-ray emission. 
Further from the nucleus are found broad-line regions, showing typical velocity widths of
a few thousand km~s$^{-1}$; reverberation mapping studies of local broad-line AGN
have yielded typical size scales of a few light-weeks for these areas. On larger
scales still are likely to lie tori of material which reprocess the quasar radiation 
and re-emit photons in the infrared, together with narrow-line emission regions a few 
hundred parsecs from the centre.

Interesting results began to emerge about a decade ago as lensed quasars were monitored
extensively at optical wavebands. It is expected that the source sizes are different
in different optical colours, because the temperature of the accretion disk increases
as its radius decreases, and this should show up as a chromatic microlensing
signal (Wambsganss \& Paczynski 1991) which was duly found in observational programmes
(e.g. Wisotzki et al. 1993, Claeskens et al. 2001, Burud et al. 2002). This effect can 
be used to estimate the accretion disk size and structure (Poindexter, Morgan \& 
Kochanek 2008, Morgan et al. 2008, Poindexter et al. 2010, Hutsemekers et 
al. 2010, Dai et al. 2010, Blackburne et al. 2011, Mu\~noz et al. 2011); the
% pmk08: HE1104 size 7E15cm, temp profile consistent with SS
% pk10: Q2237 inclination of disk, size 6e15
picture that emerges in some cases is of a scale-size of a few light days 
and a temperature-radius profile that is consistent with a standard Shakura-Sunyaev
thin disk (Poindexter et al. 2008). But this is by no means a universal result, and in
many cases the inferred size is bigger or the temperature profile is different. For 
example, Blackburne et al. (2011) analyse multiwavelength observations of a sample of 
lensed quasars and find that the microlensing properties of many of the objects imply 
accretion disk sizes of up to a factor of 10 larger than standard disks.

Comparison of the spectra of the broad emission lines in different images of quasar 
lens systems have shown that the BLR is also microlensed (Abajas et al. 2002, Richards 
et al. 2004, Wayth et al. 2005, Keeton et al. 2006, Abajas et al. 2007, Sluse et al. 
2007, Hutsemekers et al. 2010, Sluse et al. 2012) as originally predicted 30 years 
ago (Nemiroff 1988, Schneider \& Wambsganss 1990). Like the continuum microlensing 
studies, these are very important clues to the structure of the emitting object. 
Results from this work include the determination of the
overall size scale of the BLR, ranging from $<$9 light days in SDSS~J0924+0219
(Keeton et al. 2006) to a few light-months in Q2237+0305 (Wayth et al. 2005), but
lack of consensus on the structure of the BLR. This is obtained from the microlensing
signature from different velocity components within each line; in some cases there 
is evidence for an ordered, biconical structure (Abajas et al. 2007) but larger 
surveys (Sluse et al. 2012) seem to show microlensing signals which are largely 
independent in red and blue wings, suggesting a non-spherical structure for the BLR.

{\bf QL as probe of cosmology}

{\it Cosmological parameters}

Well before the discovery of the first lensed quasar, Refsdal (1964) pointed out that
a lensed quasar system could be used to measure the Hubble constant. The basic idea
is simple. Light travels along two or more different paths from the source to the
observer, via deflections at different points in the lens plane. The resulting path
difference can be measured if the background source is variable by comparing light
curves from the two images and multiplying by the speed of light. If the source and
lens redshifts are known, this then gives an absolute measure of distance together with
redshift in the system, the combination of which gives the distance scale. Fortunately, 
the expected time delays from typical lens configurations are on timescales of weeks to 
months. In
principle, this method offers a clean determination of the Hubble constant on
cosmological scales in a one-step method. Even better, in principle, measurements in
a number of different lens systems at different redshifts could also allow measurement
of $H(z)$ and hence other cosmological parameters\footnote{Early in the history of
lensing, the number of lenses in a complete sample appeared to be a useful way of
constraining $\Lambda$, because a higher $\Lambda$ increases lengths at high redshift
and hence increases the optical depth to lensing (Kochanek 1996); indeed this was the 
justification for the thoroughness of the CLASS survey in attempting to get a complete
sample of lenses. Although this line of research yielded a fairly clear result of
non-zero $\Lambda$ (Kochanek 1996, Chae et al. 2002, Mitchell et al. 2005, but see
also Keeton 2002), the rate at which constraints on dark energy improve is a very
slow function of increasing size of lens sample, and it has since been
abandoned.}.

Historically, ecstasy at the cosmological prospects after the 1979 detection of Q0957+561
quickly turned to agony, both because of the long path to the secure determination of
a time delay in Q0957+561 itself (Kundic et al. 1997), but also following the 
appreciation of the extent of the major systematic of this method.
The systematic is closely related
to the problem of determining the macromodel in a lensed system, and is that the
derived Hubble constant is effectively degenerate with the macro-properties of the
lens model, in the sense that steeper mass profiles produce lower $H_0$ for a given time
delay\footnote{Strictly, the dependence is not directly on the mass profile; it is related
to the surface mass density in the annulus between the lensed images (Kochanek 2002,
but see also Read, Saha \& Macci\`o 2007).}. 
Worse still, the mass-sheet degeneracy causes rescaling of the time delay for the same 
image positions and fluxes, and thus has an effect on $H_0$ which is unknown in a 
single-source system, unless a census of all the mass along the line of sight can be taken.

Again there are a number of responses to the problem. One is to abandon the attempt to
measure $H_0$ or cosmological parameters in general, regard them as a solved problem at 
the level that lenses will constrain, and regard time delays as a means 
to break degeneracies in mass models by using them together with a ``known'' $H_0$ -- 
for example, from the Hubble Key Project measurements of Cepheid variables. Many workers
in the field would regard this as an unnecessarily defeatist approach, given three facts:
lens $H_0$ work requires much less high-cost observing time than alternatives; large
numbers of time delays will be available in future; and although the lens modelling 
systematic is a serious systematic, it is only one systematic and not many.

The second response is to investigate a statistical approach. Can the systematic
error be reduced to a random error, albeit a large one in an individual object, which
can then be beaten down by root-$n$ statistics? This approach again begins with the
average properties of the SLACS galaxy-galaxy lenses, which appear to have a mass slope 
very close to isothermal (Koopmans et al. 2006) and mostly lie at low redshift ($<$0.3). 
More recently, a higher-redshift lens sample, known as BELLS (Brownstein et al. 2012, 
Bolton et al. 2012), has become available from a parent 
population consisting of the BOSS spectroscopic survey. These show that the mass
slope changes slightly with redshift, becoming steeper by a few tenths in the overall
power law. Matter along the line of sight is harder to control, but here again a
statistical argument may be appropriate; multiple sight-lines through large cosmological
simulations can give an indication of the possible range of amount of intervening
matter along a particular direction (although not a random direction, since lensing
is a process which takes preferentially along more crowded lines of sight). A large
Bayesian engine can then be employed to marginalise over nuisance parameters and yield
the desired information. Statistical approaches have been taken by a number of authors,
including for example Dobke et al. (2009) who calculated the number of time-delay lenses 
which would be required to constrain cosmological parameters other than $H_0$ in this
way. Similar attempts have been made to investigate $H_0$ using existing time delay
information and best-attempt mass models (Oguri 2007), yielding results around 
70$\,$kms$^{-1}$Mpc$^{-1}$ although leaving the uncomfortable feeling that mutually 
formally incompatible results for different lenses are being shoehorned into a 
harmonious conclusion. Another approach is to explore the variety of possible lens
models using non-parametric methods in order to thereby explore the possible range 
of $H_0$ for each system (Saha et al. 2006), again yielding results of 
72$\,$kms$^{-1}$Mpc$^{-1}$, with errors of about 15\%.

The third, and in my view fruitful response in the long run, is to grit one's teeth 
and do the hard work in individual cases, in the knowledge that it will get easier
as telescopes become more powerful. Two recent examples of the work required are
provided by the detailed investigations of the time-delay lenses RXJ1131$-$1231 and 
CLASS~B1608+656 by Suyu et al. (2010) and Suyu et al. (2012). A time-delay is only 
the start of such programmes. Other ingredients include deep multi-colour HST imaging, 
in order to properly model the extended structure associated with lensed extended 
emission from the quasar host and disentangle it from effects of reddening in the 
lensing galaxy; spectroscopic investigation of surrounding matter, in order to 
quantify the effects of mass sheets, taking into consideration the richness of the 
surrounding environment and comparison with cosmological simulations; radio imaging,
in the case where the lensed object is radio-loud; and performing 
the modelling blind to avoid unconscious biases, so that the conversion from an 
arbitrary scale to $H_0$ is only done once all the systematics have been estimated. 
However, when all this is done, the resulting $H_0$ values from two lenses, including 
all the systematic and random errors, are of comparable quality to the HST Key Programme 
(6\% error). There is a serious prospect from further such studies of a competitive 
contribution to the determination of $w$ which is largely orthogonal to other probes, 
once a few dozen such lenses have been thoroughly
investigated (e.g. Linder 2011).

{\it Galaxy evolution}

Individual lenses can be used to find values for the Hubble constant (or for 
ambiguity-free galaxy mass models if the cosmological world model is known). However,
the statistics of well-selected lens samples can be used to investigate galaxy evolution.
This is because the number of lenses as a function of galaxy and source redshift in
such a sample depends on the evolution in both density and mass of the available
lens population. In order to attempt this, a sample whose selection effects are 
under control is needed, and in practice the easily identifiable properties of quasars,
and the possibility of getting clean separation between lensed and unlensed objects,
makes quasar lens samples the obvious choice.

Two at least fairly complete quasar lens samples exist. The CLASS statistically 
complete survey was the result of a systematic attempt to completely identify all
lenses with separation of $>$300~mas and primary:secondary flux ratio $<$10:1. The
SQLS survey, although probably somewhat less complete at the lower-separation end,
is slightly larger. A number of authors, beginning with Chae \& Mao 2003, have
used these samples, together with a plausible cosmological world model, to derive
useful constraints on early-type galaxy evolution. Most have found a consistency with no
evolution either in number density or mass (Chae \& Mao 2003, Chae 2005, Chae, Mao \& 
Kang 2006, Matsumoto \& Futamase 2008, Chae 2010, Oguri et al. 2012), although due to the
small statistics, the error bars are still large in the index of number density
evolution. In the most recent work (Oguri et al. 2012), the evolution index of
velocity dispersion $\nu_{\sigma}\equiv d{\ln}\sigma/d\ln (1+z)$, is zero within errors
of $\sim$0.2, assuming a standard $\Lambda$-cosmology.

{\bf The future; bigger samples, and how to find them}

The current largest sample of quasars, the SDSS quasar list (Schneider et al. 2010)
contains about 100000 objects. Given a typical optical depth in lensing galaxies
towards $z\sim 2$ sources, this would imply 100-150 lensed quasars, of which at least
half have already been found in the SQLS and other surveys. In principle, many more
active galaxies are available in radio surveys - the FIRST survey, for example, contains
about 10$^6$ objects with accurate position information - but their faintness in the
radio makes systematic large surveys difficult, even with the current generation of
upgraded radio arrays. Of such arrays, the only one which has the required combination
of high sensitivity and sub-arcsecond resolution is LOFAR, a low-frequency radio 
interferometer array centred in the Netherlands (van Haarlem 2005) 
but high resolution surveys at great depth are probably a few years away.

In the optical, a combination of wide area and high resolution is likely to be achieved 
in the near future, by a number of telescopes. The first is GAIA, scheduled for 2013, 
a satellite primarily designed for astrometry and measurements of proper 
motions in Galactic stars. By virtue of area coverage and accurate measurements of
point sources, however, it is also well suited to making a large census of about
half a million quasars, to determine which such objects are extended and thus 
potentially lensed. This alone should increase the lensed quasar sample by a large factor
(Surdej et al. 2002). Towards the end of the decade, two major advances are likely 
with the advent 
of the Large Synoptic Survey Telescope (LSST) and Euclid. Euclid is an ESA medium 
mission scheduled for launch in 2019 which will have close to all-sky coverage and 
150-200 mas resolution at optical and NIR wavebands. It will provide imaging at 
slightly less angular resolution and sensitivity than the 2 square-degree COSMOS 
HST field, but over the whole sky. Its mission includes the detection of about 1000
quasar lenses, about an order of magnitude increase on the present sample (as well 
as hundreds of thousands of galaxy-galaxy lens systems). Still vaster samples will 
be provided by LSST (Abell et al. 2009), with
the additional advantage of multiple observations of the same field, allowing quasars
to be identified by variability (Kochanek et al. 2006) and probably yielding a sample of
several thousand lensed quasars. On the same timescale, the Square Kilometre Array will
provide similarly large samples, selected at radio wavelengths (Koopmans et al. 2004).

Large samples are good for two reasons. The first is that ``more of the same'' 
approaches can be tried with larger numbers, although they do rely on systematic biases 
being eliminated. If we assume that the mass slope and mass sheet model degeneracies
can be controlled, so that the accuracy improves as the square root of the number of
objects, then spectacular results can be obtained; Coe \& Moustakas (2009) calculate 
that, in conjunction with Planck priors, an accuracy of $\sim 3$\% can be obtained
on $w$. This assumes that time delays will be measurable given the LSST cadence. 
If a smaller sample of the quasar lenses are measured, however, but with more
intensive followup, then extrapolation from the work of Suyu et al. (2012) suggests
that comparable results on post-$H_0$ parameters can be achieved in conjunction with
other (BAO/SNe) cosmological probes, since the lensing constraints are often orthogonal
to others in parameter space. Linder (2011) calculates that the dark energy figure of
merit is potentially improvable by a factor of 5 by including lensing information from
future surveys.

The second advantage of large samples is that they are likely to contain a small number
of high-value objects. One of the most prized type of lens systems is a quasar lens
system with a second source at a different redshift, since such double source plane
systems allow mass model degeneracies to be immediately broken. The problem has been
investigated by Collett et al. 2012, who find that a small number of these rare 
objects give 15\% accuracy in $w$ very quickly. Many such sources would be currently
difficult or impossible to follow up, due to faintness of some of the sources, but the
era of 30-m class telescopes is around the corner, and such followup operations will become
routine if not easy.

{\bf Summary and conclusion}

The future uses of quasar lenses, as in the past, divide into three natural applications:
study of the lenses, study of the sources, and study of cosmology and galaxy evolution.
I briefly summarise the results from the main body of the review, and the prospects for
the next ten years.

\begin{itemize}
\item We already know basic facts about lens mass distributions; elliptical galaxies
have isothermal mass distributions at low redshift, and there is some indication of
steepening with redshift. In prospect is a vast increase in parameter space, with the
ability to study the evolution of galaxy mass profiles over a much wider range of
redshift, and over different masses of lens galaxies and Hubble types. Much of this
work will be done with lens-selected surveys, like the existing SLACS galaxy-galaxy
lenses; however, quasar lens systems give the opportunity to make large, source-selected
surveys and examine the statistical properties of lenses independently of their selection.
The major impact will be in the study of substructure in lens galaxies, however. The existing
sample of substructure-friendly quasar lenses is very small and has already yielded a large
and surprising body of information about the small-scale features of lensing galaxies. 
Expansion of these samples will provide critical tests for galaxy formation models.
\item Microlensing studies have yielded unprecedented insights into the nature of
quasars, particularly the proportions of stellar and smooth matter within the mass budget
and the properties -- size and physics -- of the central engine and surrounding emission
line regions. These studies require multi-wavelength observations and patient monitoring,
and have concentrated on relatively few objects. With future telescopes and high-cadence 
monitoring we can expect that studies of quasar physics at huge effective resolution will
become routine.
\item Perhaps the most important future application of quasar lensing lies in a promise
which has taken some time to fulfil, that of cosmography. Many years passed between the
discovery of the first quasar lens and the first reliable determinations of the Hubble
constant. The process is now accelerating, thanks to coordinated and long-term monitoring
campaigns, allied to advances in lens modelling and observation of individual lens systems.
Such investigations are already closing in on estimates of the Hubble constant with stringent
enough error constraints to contribute to the overall cosmological world model. Future
measurements on a large sample of quasar lenses will give cosmological parameter estimates
with error circles orthogonal to many others, and factors of several increase in figures
of merit for dark energy searches.
\item Finally, the prospect of huge lens samples carries with it the probability of finding
something totally new. We can speculate about the likelihood of possible candidates - 
completely dark lenses, cosmic strings - but with the expectation that the unexpected is
likely to prove more surprising still.
\end{itemize}

{\bf Acknowledgements}

I thank Ian Browne for a careful reading of, and comments on, the manuscript.

{\bf References}

\parskip 0mm

Abajas, C., Mediavilla, E., Mu\~noz, J.A., G\'omez-\'Alvarez, P., Gil-Merino, R., 2007, ApJ, 658, 748.
% Microlensing of a Biconical Broad-Line Region

Abajas, C., Mediavilla, E., Mu\~noz, J.A., Popovic, L.C., Oscoz, A., 2002, ApJ, 576, 640.
% The Influence of Gravitational Microlensing on the 

Abell, P.A., et al., 2009, LSST Science Book, astro-ph/0912.0201

Amara, A., Metcalf, R.B., Cox, T.J., Ostriker, J.P., 2006, MNRAS, 367, 1367.
% Simulations of strong gravitational lensing with 

Bartelmann, M., 2010, C.Q.Grav., 27 (23), 233001

Begelman, M., d.Kool, M., Sikora, M., 1991, ApJ, 382, 416.
% Outflows driven by cosmic-ray pressure in broad 

Belokurov, V., et al., 2006, ApJ, 647, L111.
% A Faint New Milky Way Satellite in Bootes

Belokurov, V., et al., 2007, ApJ, 654, 897

Biggs, A.D., et al., 2004, MNRAS, 350, 949.
% Radio, optical and infrared observations of CLASS 

Blackburne, J.A., Pooley, D., Rappaport, S., Schechter, P.L., 2011, ApJ, 729, 34.
% Sizes and Temperature Profiles of Quasar Accretion 

Blumenthal, G.R., Faber, S.M., Flores, R., Primack, J.R., 1986, ApJ, 301, 27.
% Contraction of dark matter galactic halos due to 

Bolton, A.S., Burles, S., Koopmans, L.V.E., Treu, T., Moustakas, L.A., 2006, ApJ, 638, 703.
% The Sloan Lens ACS Survey. I. A Large 

Bolton, A.S., et al., 2008, ApJ, 682, 964.
% The Sloan Lens ACS Survey. V. The Full ACS 

Bolton, A.S., et al., 2012, ApJ, 757, 82.
% The BOSS Emission-Line Lens Survey. II. 

Boyce, E., Winn, J.N., Hewitt, J.N., Myers, S.T., 2006, ApJ, 648, 73

Boylan-Kolchin, M., Bullock, J.S., Kaplinghat, M., 2011, MNRAS, 415, L40.
% Too big to fail? The puzzling darkness of massive 

Boylan-Kolchin, M., Bullock, J.S., Kaplinghat, M., 2012, MNRAS, 422, 1203
% Too big to fail? The puzzling darkness of massive 

Boylan-Kolchin, M., Springel, V., White, S.D.M., Jenkins, A., 2010, MNRAS, 406, 896.
% There's no place like home? Statistics of Milky 

Brada\v{c}, M., et al., 2004, A\&A, 423, 797.
% The signature of substructure on gravitational 

Browne, I.W.A., et al., 2003, MNRAS, 341, 13.
% The Cosmic Lens All-Sky Survey - II. Gravitational 

Brownstein, J.R., et al., 2012, ApJ, 744, 41.
% The BOSS Emission-Line Lens Survey (BELLS). I. A 

Bullock, J.S., Kravtsov, A.V., Weinberg, D.H., 2000, ApJ, 539, 517.
% Reionization and the Abundance of Galactic 

Burud, I., et al., 2002, A\&A, 383, 71.
% An optical time-delay for the lensed BAL quasar HE 

Chae, K., et al., 2002, Ph. Rev. lett., 89, 1301.
% The Cosmic Lens All-Sky Survey: statistical strong 

Chae, K., 2005, ApJ, 630, 764.
% Constraints on the Velocity Dispersion Function of 

Chae, K., 2010, MNRAS, 402, 2031.
% Galaxy evolution from strong-lensing statistics: the 

Chae, K., Mao, S., 2003, ApJ, 599, L61.
% Limits on the Evolution of Galaxies from the 

Chae, K., Mao, S., Kang, X., 2006, MNRAS, 373, 1369.
% Constraints on the velocity profiles of galaxies 

Chang, K., Refsdal, S., 1979, Natur, 282, 561.
% Flux variations of QSO 0957+561 A, B and image 

Chen, J., Koushiappas, S.M., Zentner, A.R., 2011, ApJ, 741, 117.
% The Effects of Halo-to-halo Variation on 

Chen, J., et al., 2007, ApJ, 659, 52

Chiba, M., 2002, ApJ, 565, 17.
% Probing Dark Matter Substructure in Lens Galaxies

Chiba, M., Minezaki, T., Kashikawa, N., Kataza, H., Inoue, K.T., 2005, ApJ, 627, 53.
% Subaru Mid-Infrared Imaging of the Quadruple Lenses 

Chwolson, O., 1924, AN, 221, 329.
%* Über eine mögliche Form fiktiver Doppelsterne

Claeskens, J., Surdej, J., 2002, A\&ARv, 10, 263.
%* Gravitational lensing in quasar samples

Claeskens, J.-F., Khmil, S.V., Lee, D.W., Sluse, D., Surdej, J., 2001, A\&A, 367, 748.
% HST and ground-based observations of the 

Coe, D., Moustakas, L.A., 2009, ApJ, 706, 45.
% Cosmological Constraints from Gravitational Lens 

Cohn, J.D., Kochanek, C.S., McLeod, B.A., Keeton, C.R., 2001, ApJ, 554, 1216.
% Constraints on Galaxy Density Profiles from Strong 

Collett, T.E., Auger, M.W., Belokurov, V., Marshall, P.J., Hall, A.C., 2012, MNRAS, 424, 2864.
% Constraining the dark energy equation of state with 

Congdon, A.B., Keeton, C.R., Nordgren, C.E., 2008, MNRAS, 389, 398.
% Analytic relations for magnifications and time 

Courbin, F., Saha, P., Schechter, P.L., 2002, LNP, 608, 1.
%* Quasar Lensing

Croton, D.J., et al., 2006, MNRAS, 365, 11.
% The many lives of active galactic nuclei: cooling 

Dai, X., et al., 2010, ApJ, 709, 278.
% The Sizes of the X-ray and Optical Emission Regions 

Dalal, N., Kochanek, C.S., 2002, ApJ, 572, 25.
% Direct Detection of Cold Dark Matter Substructure

Dobke, B., et al., 2009, MNRAS, 397, 311

Dobler, G., Keeton, C.R., 2006, MNRAS, 365, 1243.
% Finite source effects in strong lensing: 

El\'{\i}asd\'ottir A.,Hjorth J., Toft S., Burud, I., Paraficz, D., 2006, ApJS, 166, 443

Faber, S.M., et al., 1997, AJ, 114, 1771.
% The Centers of Early-Type Galaxies with HST. IV. 

Fadely, R., Keeton, C.R., 2011, AJ, 141, 101.
% Near-infrared K and L' Flux Ratios in Six Lensed 

Fadely, R., Keeton, C.R., 2012, MNRAS, 419, 936.
% Substructure in the lens HE 0435-1223

Falco, E.E., Gorenstein, M.V., Shapiro, I.I., 1985, ApJ, 289, L1.
% On model-dependent bounds on H(0) from gravitational 

Fassnacht, C.D., et al., 1999, AJ, 117, 658.
% B2045+265: A New Four-Image Gravitational Lens from 

% Fassnacht, C.D., et al., 1999, ApJ, 527, 498.
% A Determination of H<SUB>0</SUB> with the CLASS Gravitational 

Fassnacht, C.D., et al., 2006, ApJ, 642, 30
% stuff along LOS

Gnedin, O.Y., Kravtsov, A.V., Klypin, A.A., Nagai, D., 2004, ApJ, 616, 16.
% Response of Dark Matter Halos to Condensation of 

Gorenstein, M.V., Shapiro, I.I., Falco, E.E., 1988, ApJ, 327, 693.
% Degeneracies in parameter estimates for models of 

van Haarlem M., 2005, EAS 15, 431

Haarsma, D.B., et al., 2005, AJ, 130, 1977.
% The FIRST-Optical-VLA Survey for Lensed Radio Lobes

Heckman, T.M., Armus, L., Miley, G.K., 1990, ApJS, 74, 833.
% On the nature and implications of starburst-driven 

Hewitt, J.N., Turner, E.L., Schneider, D.P., Burke, B.F., Langston, G.I., 1988, Natur, 333, 537.
%* Unusual radio source MG1131+0456 - A possible 

Huchra, J., Gorenstein, M., Kent, S., Shapiro, I., Smith, G., Horine, E., Perley, R.,
1985, AJ, 90, 691

Hutsem\'ekers, D., Borguet, B., Sluse, D., Riaud, P., Anguita, T., 2010, A\&A, 519A, 103.
% Microlensing in H1413+117: disentangling line 

Inada, N., et al., 2003, Natur, 426, 810.
% A gravitationally lensed quasar with quadruple 

Inada, N., et al., 2005, PASJ, 57, L7

Inada, N., et al., 2006, ApJ, 653, L97.
% SDSS J1029+2623: A Gravitationally Lensed Quasar 

Inada, N., et al., 2008, AJ, 135, 496.
% The Sloan Digital Sky Survey Quasar Lens Search. II. 

Inada, N., et al., 2010, AJ, 140, 403.
% The Sloan Digital Sky Survey Quasar Lens Search. IV. 

Inada, N., et al., 2012, AJ, 143, 119.
% The Sloan Digital Sky Survey Quasar Lens Search. V. 

Inoue, K.T., Takahashi, R., 2012, MNRAS, 426, 2978.
% Weak lensing by line-of-sight haloes as the origin 

Irwin, M.J., Webster, R.L., Hewett, P.C., Corrigan, R.T., Jedrzejewski, R.I., 1989, AJ, 98, 1989.
% Photometric variations in the Q2237 + 0305 system - 

Jackson, N., 2007, LRR, 10, 4.
%* The Hubble Constant

Jackson, N., 2011, ApJ, 739, L28.
% The Faintest Radio Source Yet: Expanded Very Large 

Jackson, N., Rampadarath, H., Ofek, E.O., Oguri, M., Shin, M., 2012, MNRAS, 419, 2014
%* Muscles paper

Jackson, N., Ofek, E.O., Oguri, M., 2009, MNRAS, 398, 1423.
% A new gravitational lens from the MUSCLES survey: 

Kayser, R., 1986, A\&A, 157, 204.
% Gravitational lenses - Model fitting, time delay and 

Kayser, R., Refsdal, S., 1989, Natur, 338, 745.
% Detectability of gravitational microlensing in the 

Kayser, R., Refsdal, S., Stabell, R., 1986, A\&A, 166, 36.
% Astrophysical applications of gravitational 

Kayser, R., Refsdal, S., Weiss, A., Schneider, P., 1989, A\&A, 214, 4.
% Gravitational micro-lensing due to an ensemble of 

Keeton, C.R., 2002, ApJ, 575, L1

Keeton, C.R., 2003, ApJ, 582, 17.
% Lensing and the Centers of Distant Early-Type 

Keeton, C.R., Burles, S., Schechter, P.L., Wambsganss, J., 2006, ApJ, 639, 1.
% Differential Microlensing of the Continuum and Broad 

Keeton, C.R., Gaudi, B.S., Petters, A.O., 2003, ApJ, 598, 138.
% Identifying Lenses with Small-Scale Structure. I. 

Keeton, C.R., Gaudi, B.S., Petters, A.O., 2005, ApJ, 635, 35.
% Identifying Lenses with Small-Scale Structure. II. 

Keeton, C.R., Moustakas, L.A., 2009, ApJ, 699, 1720.
% A New Channel for Detecting Dark Matter Substructure 

Keeton, C.R., Zabludoff, A.I., 2004, ApJ, 612, 660.
% The Importance of Lens Galaxy Environments

Klypin, A., Kravtsov, A.V., Valenzuela, O., Prada, F., 1999, ApJ, 522, 82.
% Where Are the Missing Galactic Satellites?

Kochanek, C.S., 1996, ApJ, 466, 638.
% Is There a Cosmological Constant?

Kochanek, C.S., 2002, ApJ, 578, 25.
% What Do Gravitational Lens Time Delays Measure?

Kochanek, C.S., 2004, ApJ, 605, 58.
% Quantitative Interpretation of Quasar Microlensing 

Kochanek, C.S., Dalal, N., 2004, ApJ, 610, 69.
% Tests for Substructure in Gravitational Lenses

Kochanek, C.S., Mochejska, B., Morgan, N.D., Stanek, K.Z., 2006, ApJ, 637, L73.
% A Simple Method to Find All Lensed Quasars

Kochanek, C.S., Schechter, P.L., 2004, Measuring and Modeling the Universe, 
from the Carnegie Observatories Centennial Symposia. Published by Cambridge 
University Press, as part of the Carnegie Observatories Astrophysics Series. 
Ed. W.L. Freedman, p. 117.
%* The Hubble Constant from Gravitational Lens Time 

Kochanek, C.S., 2004. In Kochanek, C.S., Schneider, P., Wambsganss, J., 2004, 
Gravitational Lensing: Strong, Weak \& Micro, Proceedings of the 33rd Saas-Fee 
Advanced Course, G. Meylan, P. Jetzer \& P. North, eds. (Springer-Verlag: Berlin)
%* Saas-Fee

Kochanek, C.S., White, M., 2001, ApJ, 559, 531.
% Global Probes of the Impact of Baryons on Dark 

Kochanek, C.S., et al., 2006, ApJ, 640, 47.
% The Time Delays of Gravitational Lens HE 0435-1223: 

Koopmans, L.V.E., Browne, I.W.A., Jackson, N.J., 2004, NewAR, 48, 1085.

Koopmans, L.V.E., Treu, T., Bolton, A.S., Burles, S., Moustakas, L.A., 2006, ApJ, 649, 599.
% The Sloan Lens ACS Survey. III. The Structure and 

Koopmans, L.V.E., et al., 2003, ApJ, 595, 712

Kratzer, R.M., et al., 2011, ApJ, 728, L18.
% Analyzing the Flux Anomalies of the Large-separation 

Kundic, T., et al., 1997, ApJ, 482, 75.
% A Robust Determination of the Time Delay in 

Lagattuta, D.J., Auger, M.W., Fassnacht, C.D., 2010, ApJ, 716, L185.
% Adaptive Optics Observations of B0128+437: A 

Lawrence, A., et al., 2007, MNRAS, 379, 1599.
% The UKIRT Infrared Deep Sky Survey (UKIDSS)

Leh\'ar, J., Buchalter, A., McMahon, R.G., Kochanek, C.S., Muxlow, T.W.B., 2001, ApJ, 547, 60.
% An Efficient Search for Gravitationally Lensed Radio 

Linder, E.V., 2011, PhRvD, 84l3529L, .
% Lensing time delays and cosmological complementarity

MacLeod, C.L, Jones, R., Agol, E., Kochanek, C.S., 2012, astro-ph/1212.2166
% Detection of Substructure in the Gravitationally 

MacLeod, C.L., Kochanek, C.S., Agol, E., 2009, ApJ, 703, 1177

Macci\'o, A.V., Moore, B., Stadel, J., Diemand, J., 2006, MNRAS, 366, 1529.

Mao, S., Witt, J., Koopmans, L.V.E., 2001, MNRAS, 323, 301

Mao, S., Jing, Y., Ostriker, J.P., Weller, J., 2004, ApJ, 604, L5.
% Anomalous Flux Ratios in Gravitational Lenses: For 

Mao, S., Schneider, P., 1998, MNRAS, 295, 587.
% Evidence for substructure in lens galaxies?

Matsumoto, A., Futamase, T., 2008, MNRAS, 384, 843.
% Validity of strong lensing statistics for 

McKean, J.P., et al., 2007, MNRAS, 378, 109.
% High-resolution imaging of the anomalous flux ratio 

Metcalf, R.B., 2002, ApJ, 580, 696.
% The Detection of Pure Dark Matter Objects with Bent 

Metcalf, R.B., 2005, ApJ, 622, 72.
% Testing LambdaCDM with Gravitational Lensing Constraints 

Metcalf, R.B., 2005, ApJ, 629, 673.
% The Importance of Intergalactic Structure to 

Metcalf, R.B., Amara, A., 2012, MNRAS, 419, 3414.
% Small-scale structures of dark matter and flux 

Metcalf, R.B., Zhao, H., 2002, ApJ, 567, L5.
% Flux Ratios as a Probe of Dark Substructures in 

Miranda, M., Macci\`o, A.V., 2007, MNRAS, 382, 1225.
% Constraining warm dark matter using QSO 

Mitchell, J.L., Keeton, C.R., Frieman, J.A., Sheth, R.K., 2005, ApJ, 622, 81.
% Improved Cosmological Constraints from Gravitational 

Momcheva, I., Williams, K., Keeton, C., Zabludoff, A., 2006, ApJ, 641, 169.
% environments of lens galaxies

Moore, B., et al., 1999, ApJ, 524, L19.
% Dark Matter Substructure within Galactic Halos

Morgan, C.W., Kochanek, C.S., Dai, X., Morgan, N.D., Falco, E.E., 2008, ApJ, 689, 755.

Moustakas, L.A., et al., 2012, AAS, 219, 146.01.
% The Orphan Lenses Project

Mu\~noz, J.A., Mediavilla, E., Kochanek, C.S., Falco, E.E., Mosquera, A.M., 2011, ApJ, 742, 67.
% A Study of Gravitational Lens Chromaticity with the 

Myers, S.T., et al., 1999, AJ, 117, 2565.
% CLASS B1152+199 and B1359+154: Two New Gravitational 

Myers, S.T., et al., 2003, MNRAS, 341, 1.
% The Cosmic Lens All-Sky Survey - I. Source selection 

Nemiroff, R.J., 1988, ApJ, 335, 593.
% AGN broad emission line amplification from 

Oguri, M., 2007, ApJ, 660, 1.
% Gravitational Lens Time Delays: A Statistical 

Oguri, M., et al., 2006, AJ, 132, 999.
% The Sloan Digital Sky Survey Quasar Lens Search. I. 

Oguri, M., et al., 2012, AJ, 143, 120.
% The Sloan Digital Sky Survey Quasar Lens Search. VI. 

Paczynski, B., 1986, ApJ, 301, 503.
% Gravitational microlensing at large optical depth

Phillips, P.M., et al., 2000, MNRAS, 319, L7.
% A new quadruple gravitational lens system: CLASS 

Poindexter, S., Kochanek, C.S., 2010, ApJ, 712, 668.
% Microlensing Evidence that a Type 1 Quasar is Viewed 

Poindexter, S., Morgan, N., Kochanek, C.S., 2008, ApJ, 673, 34.
% The Spatial Structure of an Accretion Disk

Read, J.I., Saha, P., Macci\`o, A.V., 2007, ApJ, 667, 645.
% Radial Density Profiles of Time-Delay Lensing 

Refsdal, S., 1964, MNRAS, 128, 307.
% On the possibility of determining Hubble's parameter 

Richards, G.T., et al., 2004, ApJ, 610, 679.
% Microlensing of the Broad Emission Line Region in 

Rickett, B.J., 1977, ARA\&A, 15, 479.
% Interstellar scattering and scintillation of radio 

Ros, E., et al., 2000, A\&A, 362, 845.
% VLBI imaging of the gravitational lens MG J0414+0534

Rusin, D., Ma, C., 2001, ApJ, 549, L33.
% Constraints on the Inner Mass Profiles of Lensing 

Rusin, D., et al., 2002, MNRAS, 330, 205.
% High-resolution observations and mass modelling of 

Rusin, D., et al., 2003, ApJ, 587, 143.

Ryden, B., 1988, MNRAS, 329, 589.

Saha, P., 2000, AJ, 120, 1654.
% Lensing Degeneracies Revisited

Saha, P., Williams, L.L.R., Ferreras, I., 2007, ApJ, 663, 29.
% Meso-Structure in Three Strong-lensing Systems

Saha, P., Coles, J., Macci\`o, A., Williams, L., 2006, ApJ, 650, L17.

Schechter, P., Moore, C.B., 1993, AJ, 105, 1.

Schmidt, R., Wambsganss, J., 1998, A\&A, 339, 397.
% Limits on MACHOs in the lensing galaxy 0957+561

Schmidt, R.W., Wambsganss, J., 2010, Gen.Rel.Grav., 42, 2127.
%* Quasar microlensing

Schneider, P., Ehlers, J., Falco, E.E., 1992, ``Gravitational Lenses''. Springer, Berlin.
%* Gravitational Lenses

Schneider, P., Wambsganss, J., 1990, A\&A, 237, 42.
% Are the broad emission lines of quasars affected by 

Schneider, P., Weiss, A., 1992, A\&A, 260, 1.
% The gravitational lens equation near cusps

Schneider, D., et al., 2010, astro-ph/1004.1167.

Shin, E.M., Evans, N.W., 2008, MNRAS, 385, 2107.
% The effect of satellite galaxies on gravitational 

Sluse, D., Claeskens, J.-F., Hutsem\'ekers, D., Surdej, J., 2007, A\&A, 468, 885.
% Multi-wavelength study of the gravitational lens 

Sluse, D., Hutsem\'ekers, D., Courbin, F., Meylan, G., Wambsganss, J., 2012, A\&A, 544A, 62.
% Microlensing of the broad line region in 17 lensed 

Soldner, J., 1801. Berl. Astr. Jahrbuch., 1804, 161
%* defl light

Surdej J., Claeskens J.-F., Smette A., 2002, GAIA WG meeting, IAGL,

Suyu, S.H., 2012, astro-ph/1208.6010
% Cosmography from two-image lens systems: overcoming 

Suyu, S.H., et al., 2010, ApJ, 711, 201.
% Dissecting the Gravitational lens B1608+656. II. 

Vegetti, S., Koopmans, L.V.E., Bolton, A., Treu, T., Gavazzi, R., 2010, MNRAS, 408, 1969.
% Detection of a dark substructure through 

Vegetti, S., et al., 2012, Natur, 481, 341.
% Gravitational detection of a low-mass dark satellite 

Wallington, S., Narayan, R., 1993, ApJ, 403, 517

Walsh, D., Carswell, R.F., Weymann, R.J., 1979, Natur, 279, 381.
%* 0957 + 561 A, B - Twin quasistellar objects or 

Wambsganss, J., 1998, LRR, 1, 12
%* Gravitational lensing in astronomy

Wambsganss, J., 2004. In Kochanek, C.S., Schneider, P., Wambsganss, J., 2004, 
Gravitational Lensing: Strong, Weak \& Micro, Proceedings of the 33rd Saas-Fee 
Advanced Course, G. Meylan, P. Jetzer \& P. North, eds. (Springer-Verlag: Berlin)
%* Saas-fee

Wambsganss, J., Paczynski, B., 1991, AJ, 102, 864.
% Expected color variations of the gravitationally 

Wayth, R.B., O'Dowd, M., Webster, R.L., 2005, MNRAS, 359, 561.
% A microlensing measurement of the size of the broad 

White, S.D.M., Rees, M.J., 1978, MNRAS, 183, 341

White, R.L., Helfand, D.J., Becker, R.H., Glikman, E., de Vries, W.,
2007, ApJ, 654, 99

Winn, J.N., et al., 2001, AJ, 121, 1223
%* discovery paper of a southern-class lens

Winn, J.N., et al., 2002, AJ, 123, 10.
% PMN J1632-0033: A New Gravitationally Lensed Quasar 

Winn, J.N., Rusin, D., Kochanek, C.S., 2003, ApJ, 587, 80

Winn, J.N., Rusin, D., Kochanek, C.S., 2004, Nat, 427, 613

Wisotzki, L., et al., 2003, A\&A, 408, 455

Wisotzki, L., Koehler, T., Kayser, R., Reimers, D., 1993, A\&A, 278, L15.
% The new double QSO HE 1104-1805: Gravitational lens 

Wucknitz, O., 2002, MNRAS, 332, 951.
% Degeneracies and scaling relations in general 

Wyithe, J.S.B., Webster, R.L., Turner, E.L., Mortlock, D.J., 2000, MNRAS, 315, 62.
% A gravitational microlensing determination of 

Xu, D.D., et al., 2009, MNRAS, 398, 1235.
% Effects of dark matter substructures on 

Xu, D.D., et al., 2012, MNRAS, 421, 2553.
% On the effects of line-of-sight structures on 

Yonehara, A., 2001, ApJ, 548, L127.
% Evidence for a Source Size of Less than 2000 AU in 

York, D.G., et al., 2000, AJ, 120, 1579.
% The Sloan Digital Sky Survey: Technical Summary

Zackrisson, E., Riehm, T., 2010, Adv. Ast., 478910, astro-ph/0905.4075 
% Gravitational Lensing as a Probe of Cold Dark Matter 

Zackrisson, E., et al., 2012, astro-ph/1208.5482

Zhang, M., Jackson, N., Porcas, R.W., Browne, I.W.A., 2007, MNRAS, 377, 1623.
% A search for the third lensed image in JVAS 

Zucker, D.B., et al., 2006a, ApJ, 643L, 103.
% A New Milky Way Dwarf Satellite in Canes Venatici

Zucker, D.B., et al., 2006b, ApJ, 650L, 41.
% A Curious Milky Way Satellite in Ursa Major

\end{document}